\documentclass[12pt]{iopart}

\usepackage{graphicx}
\usepackage{iopams}  

\begin{document}

\title{Probing of negatively charged and neutral excitons in monolayer MoS$_{2}$}

\author{J Jadczak\textit{$^{1}$},
 J Kutrowska-Girzycka\textit{$^{1}$},
 M Bieniek\textit{$^{2,3}$},
 T Kazimierczuk\textit{$^{4}$}, 
 P Kossacki\textit{$^{4}$}, 
 J J Schindler\textit{$^{5}$},
 J Debus\textit{$^{5}$},
 K Watanabe\textit{$^{6}$}, 
 T Taniguchi\textit{$^{6}$}, 
 C H Ho\textit{$^{7}$}, 
 A W\'ojs\textit{$^{2}$}, 
 P \ Hawrylak\textit{$^{3}$} and 
 L Bryja\textit{$^{1}$}}
 
\address{\textit{$^{1}$~Department of Experimental Physics, Wroc\l aw University of Science and Technology, Wybrze\.ze Wyspia\'nskiego 27, 50-370 Wroc\l aw, Poland}}

\address{\textit{$^{2}$~Department of Theoretical Physics, Wroc\l aw University of Science and Technology, Wybrze\.ze Wyspia\'nskiego 27, 50-370 Wroc\l aw, Poland }}

\address{\textit{$^{3}$~Department of Physics, University of Ottawa, Ottawa, Ontario, Canada K1N 6N5}}

\address{\textit{$^{4}$~Institute of Experimental Physics, Faculty of Physics, University of Warsaw, Pasteura 5, 02-093 Warsaw, Poland }}

\address{\textit{$^{5}$~Experimentelle Physik 2, Technische Universität Dortmund, 44227 Dortmund, Germany }}

\address{\textit{$^{6}$~National Institute for Materials Science, Tsukuba, Ibaraki, 305-0044, Japan }}

\address{\textit{$^{7}$~Department of Electronic Engineering, National Taiwan University of Science and Technology, Taipei, 106, Taiwan}}

\ead{joanna.jadczak@pwr.edu.pl}
\vspace{10pt}
\begin{indented}
\item[]March 2020
\end{indented}

\begin{abstract}
We present results of optical experiments and theoretical analysis on the high-quality single-layer MoS$_{2}$ which reveal the fine structure of charged excitons, i.e., trions. In the emission spectra we resolve and identify two trion peaks, T$_{1}$ and T$_{2}$, resembling the pair of singlet and triplet trion peaks (T$_S$ and T$_{T}$) in tungsten-based materials. However, in polarization-dependent photoluminescence measurements we identify these peaks as intra- and inter-valley singlet trions due to the trion fine structure distinct from that already known in bright and dark 2D materials with large conduction-band splitting induced by the spin-orbit coupling. We show that the trion energy splitting in MoS$_{2}$ is a sensitive probe of inter- and intra-valley carrier interaction. With additional support from theory we claim that the existence of these singlet trions combined with an anomalous excitonic g-factor together suggest that monolayer MoS$_{2}$ has a dark excitonic ground state, despite having "bright" single-particle arrangement of spin-polarized conduction bands.
\end{abstract}

%
\vspace{2pc}

\section{Introduction} 

Monolayer transition metal dichalcogenides (TMDs) have attracted considerable scientific interest due to their unique physical properties \cite{Geim-Grigorieva-2013,Mak-Shan-2018,Wang-Urbaszek-2018}, such as direct optical bandgaps located at $\pm K$ valleys of two dimensional (2D) hexagonal Brillouin zone \cite{Mak-Heinz-2010} and coupling of spin and valley degrees of freedom \cite{Xiao-Yao-2012}. Their 2D nature and reduced dielectric screening enable formation of excitons with binding energies of few hundreds of meV \cite{Qiu-Louie-2013}, orders of magnitude larger than those in typical quasi-2D quantum wells \cite{Dingle-Henry-1974, Jadczak-2012}. In addition to neutral excitons, in the presence of excess carriers, also trions can be formed \cite{Lampert-1958, Kheng-Cox-1993, Hawrylak-1991, Bryja-2012, Jadczak-2012a, Narvaez-Hawrylak-Brum-2001} with binding energies ($E_{B}$) of tens of meV, resulting in their stability even at room temperatures \cite{Mak-Shan-2013, Jadczak-Bryja-2017,Jadczak_robust_2017}. Consequently, the monolayer TMDs constitute a novel platform to optically probe many-body effects in the presence of strong Coulomb interactions \cite{Hawrylak-1991,Hawrylak-1990,  Brum-Brown-1996, Tuan-Dery-2017, Tuan-Dery-2019, Jadczak-2019}, particularly when the Fermi level in  the conduction band for n-doped samples is lower or of the same order as the trion binding energy \cite{Drupppel-Rohlfing-2017, Scrace-2015, Roch-Warburton-2019}.

Monolayer MoS$_2$ is probably the best-known member of semiconducting TMDs \cite{Wilson-Yoffe-1969} due to its natural abundance in almost chemically pure mineral molybdenite \cite{Anthony-Nichols-2003}. Compared to other materials from the MX$_{2}$ (M = Mo, W; X = S, Se, Te) family, MoS$_{2}$ is unique having a small spin splitting ($\sim 3$ meV) of the conduction band  \cite{Kadantsev-Hawrylak-2012}, in comparison to typical Fermi energies in n-type exfoliated TMDs monolayers (ML) transferred onto standard SiO$_2$/Si substrate \cite{Wang-Urbaszek-2018,Radisavljevic-Kis-2011}. Until recently, the optical quality of monolayer MoS$_2$ was inferior in quality to the related materials, such as MoSe$_2$, WS$_2$, and WSe$_2$. This allowed for observation of only broad photoluminescence peaks dominated by trions \cite{Mak-Shan-2013}. However, progress was made by encapsulation of a MoS$_2$ monolayer with hexagonal boron nitride (hBN) layers, resulting in narrow neutral exciton lines with a width of about 2 meV at low temperatures (4 K) \cite{Cadiz-Urbaszek-2017}. The high-quality MoS$_2$/hBN van der Waals heterostructures hence enabled the observation of subtle optical and spin-valley properties of monolayer MoS$_2$ \cite{Cadiz-Urbaszek-2017, Goryca-Crooker-2019, Roch-Warburton-2019}. 

Recent works reported of a trion emission line splitting in the presence of carriers in gated nanostructures \cite{Roch-Warburton-2019}, similar to other "dark" TMDs \cite{Lyons-Tartakovskii-2019}. Within a trion model, inter- and intra- valley singlet and triplet trion nature of these split lines was suggested \cite{Drupppel-Rohlfing-2017}, with a splitting energy of about 5 meV. Contrary to those calculations, triplet trion in MoS$_{2}$ was shown to be unbound in reference \cite{Tempelaar-Berkelbach-2019}, predicting intra-/inter- valley singlet-singlet splitting $\sim 4$ meV.  Interestingly, small values of an effective exciton g-factor in hBN encapsulated MoS$_{2}$ monolayers, $g=-1.7$ in magneto-photoluminescence experiments \cite{Cadiz-Urbaszek-2017} and $g=-3.0$ in magneto-transmission measurements \cite{Goryca-Crooker-2019}, have also been observed. Furthermore, detailed magneto-transmission experiments have revealed that the reduced mass of the exciton in monolayer MoS$_2$ is heavier than predicted by density functional theories \cite{Kadantsev-Hawrylak-2012, Kormanyos-Falko-2015}, suggesting a large electron mass consistent with recent values determined from transport studies of n-type MoS$_{2}$ monolayers \cite{Pisoni-Ensslin-2018}. The complicated interplay of band structure \cite{Kadantsev-Hawrylak-2012, Bieniek_Hawrylak_2018}, electron-electron interactions \cite{Hawrylak-1991, Scrace-2015, Efimkin-MacDonald-2017, Gao-Yang-2017}, enhanced spin-splitting \cite{Ferreiros-Cortijo-2014}, dynamical effects \cite{Liu-Zhu-2019}  and inter-/ intra- valley phonon \cite{Glazov-Marie-2019} and plasmon \cite{Tuan-Dery-2017} contributions makes MoS$_{2}$ one of the most challenging materials to understand.

\begin{figure}[h]
    \centering
    \includegraphics[width=0.45\textwidth]{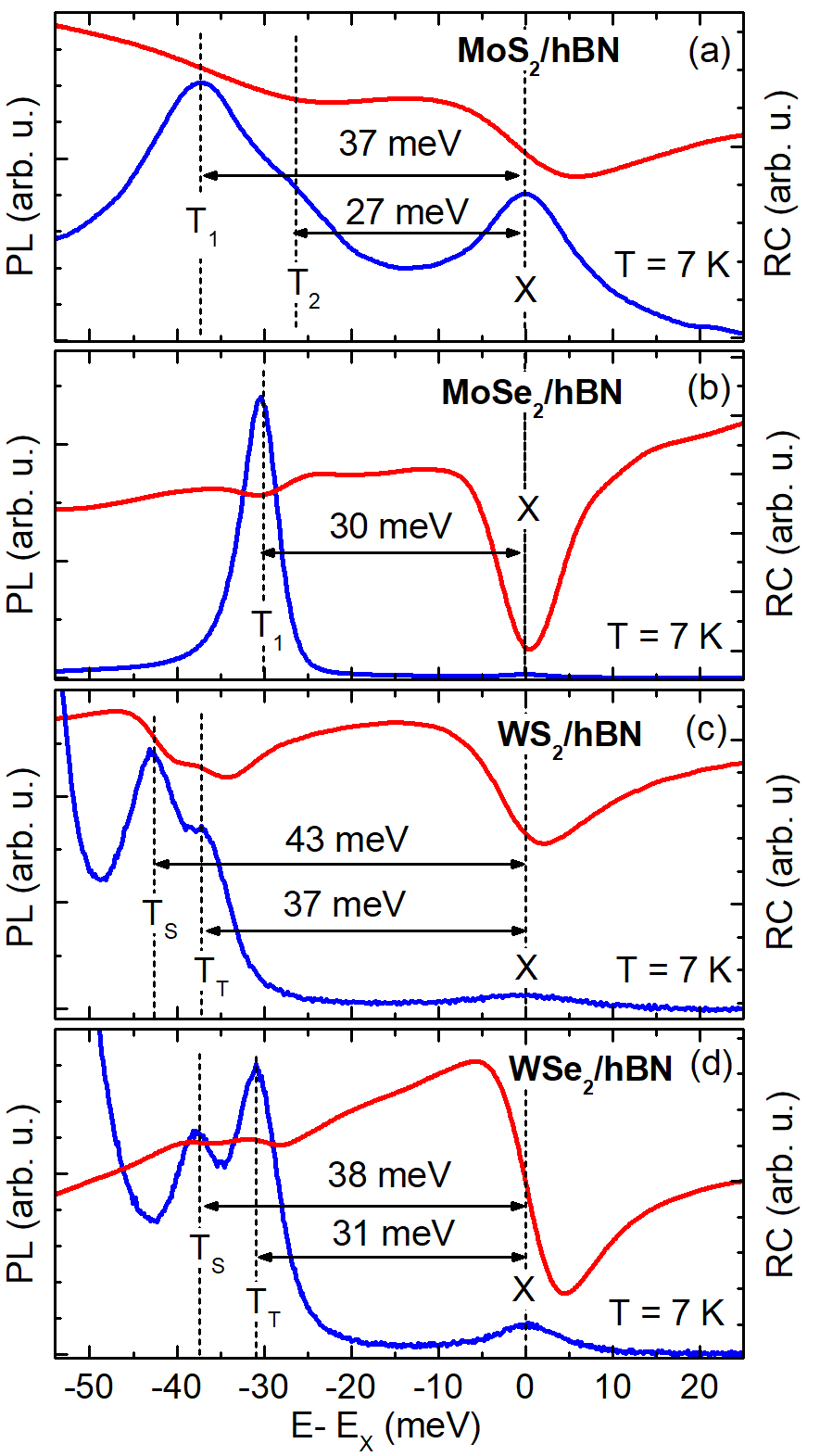}
    \caption{The comparative PL and RC spectra for (a) MoS$_{2}$/hBN, (b) MoSe$_{2}$/hBN, (c) WS$_{2}$/hBN, (d) WSe$_{2}$/hBN structures measured at 7 K.} \label{fig1}
\end{figure}

\section{PL and RC measurements of MX$_{2}$} 
As a first step toward this understanding we present  results of photoluminescence (PL) and reflectance contrast (RC) measurements of Mo- and W-based monolayer TMDs. Figure \ref{fig1}(a-d) compares low temperature (7 K) PL and RC spectra of molybdenum and tungsten based TMDs monolayers deposited on hBN/SiO$_2$/Si substrate. The PL spectra are excited non-resonantly at an energy of 2.33 eV. For all monolayers, the optical transitions are associated with the nearly free states of the neutral exciton (X) \cite{Jadczak-Bryja-2017} and different types of trions (T$_1$, T$_2$, T$_S$, T$_T$) \cite{Courtade-Urbaszek-2017, Vaclavkova-Molas-2018}. In all spectra the energies are measured from exciton energy. In figure \ref{fig1}(a) we show PL and RC spectra for MoS$_2$ while figure \ref{fig1}(b) shows the same spectra for MoSe$_2$. Interestingly, in the PL spectrum of MoS$_2$ we see an exciton peak X and two, T$_1$ and T$_2$, peaks at lower energy. As indicated in figure \ref{fig1}(a) the energy separation between these optical trion transitions is $\Delta\approx 10$ meV. We attribute them tentatively to the recombination from trion states. The MoS$_2$  spectra are to be contrasted with the spectra of MoSe$_2$, see figure \ref{fig1}(b), which demonstrate an exciton and only a single trion T$_1$ line. However, the doublet structure of the trion emission line positioned below the neutral exciton in MoS$_2$ is similar to the trion emission spectra in tungsten based monolayers, shown in figure \ref{fig1}(c-d). In the PL and RC spectra of WS$_2$ (figure \ref{fig1}(c)) and WSe$_2$ (figure \ref{fig1}(d)), the T$_{S}$ and T$_T$ optical transitions are identified at low doping level ($E_F<<E_B$). The energy difference between T$_S$ and T$_T$ is equal to $\sim 6$ meV, and is in good agreement with the recently reported values \cite{Vaclavkova-Molas-2018}. As seen in figures \ref{fig1}(c) and \ref{fig1}(d), the T$_T$ emission line in the PL spectra of WSe$_2$ is more prominent than the corresponding optical transition observed in the PL spectra of WS$_2$. This feature likely depends on the different two-dimensional electron gas (2DEG) concentration in the studied monolayers, which in sulfides is typically two orders of magnitude higher than in selenides \cite{Radisavljevic-Kis-2011, Ross-Xu-2013}.

\begin{figure}[h]
    \centering
    \includegraphics[width=0.45\textwidth]{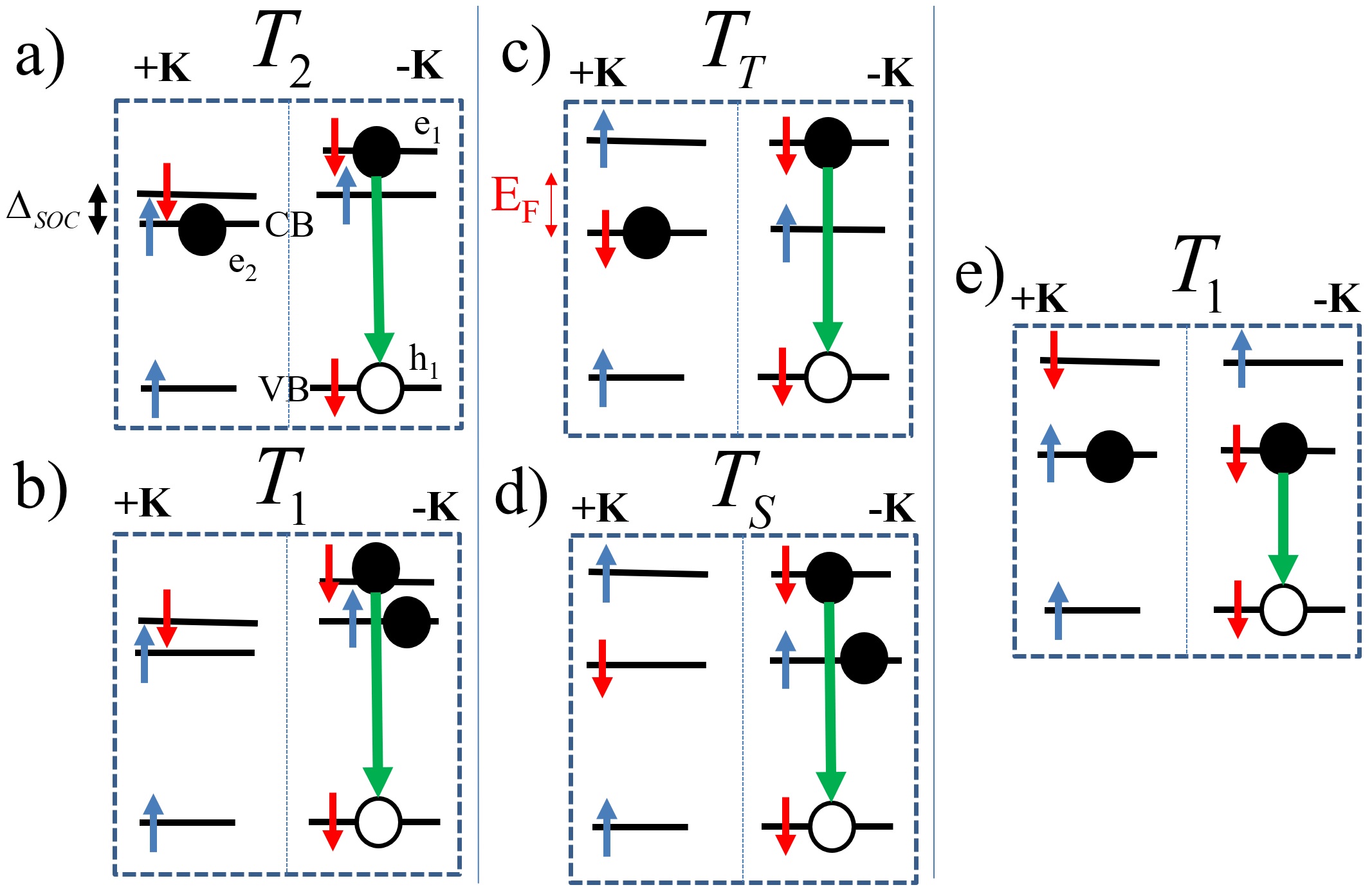}
    \caption{Summary of optically bright trion possibilities in MX$_2$. (a) Inter- and (b) intra- valley singlet-singlet trions in material with small CB spin splitting and dark exciton ground state. (c) Inter- and (d) intra- valley triplet-singlet trions in material with dark bands and dark exciton ground state. (e) Trion in compounds with bright bands arrangement and bright exciton ground state. Left panels on (a)-(e) show single particle states in the $+K$ valley, the right ones - electron configurations including electron-electron interactions.} 
		\label{fig2}
\end{figure}

\section{Theoretical considerations} 
Our interpretation of the two emission lines resolved in the spectra of MoS$_2$, shown in figure \ref{fig1}(a), as originating from two trion states, T$_1$ and T$_2$, is summarized in figures \ref{fig2}(a-b). Figure \ref{fig2}(a) shows schematically an electronic configuration for the inter-valley singlet trion T$_2$, with one spin-up electron in the $+K$ valley, a second electron with spin-down and a missing spin-down valence-band electron (valence hole) in the $-K$ valley. The left column with single particle states in the $+K$ valley considers a positive spin-orbit splitting of the conduction band states and optically active single particle transitions. The right column shows electron configurations including electron-electron interactions. It is worthwhile to mention that in this picture the energy of the spin-down $-K$ valley electron is above the spin-up empty electron state, thus, the order of the electron states is inverted and the lowest energy state is dark. We now explain \cite{Bieniek-Hawrylak-2020} this bright-dark ground excitonic state inversion shown in figure \ref{fig2}(a). The first mechanism is related to different masses of spin-up and -down electrons in the conduction band. The higher energy spin-up conduction band in the $-K$ valley is heavier; hence the higher electron mass combined with electron-hole attraction results in a more strongly bound dark state, pushing the electron-hole bright configuration up in energy. The second contribution adding to the blue shift of the energy of the bright configuration is the repulsive electron-hole exchange interaction. The two effects result in a splitting of the bright-dark 1s excitons up to 9 meV. Figure \ref{fig2}(b) shows the configuration of the intra-valley trion where two electrons in the CB are in the same $-K$ valley and hence necessarily in the singlet state. Accordingly, both trions T$_1$ and T$_2$ are bright singlet trions and the lowest energy trion is dark.

The unusual arrangement of the trion states in a material with a positive and small spin-orbit induced splitting of the conduction band is contrary to trion states in materials with a negative and large spin-orbit induced splitting of the conduction band such as WS$_2$ and WSe$_2$, see figure \ref{fig2}(c,d). The single particle arrangement of levels in the $+K$ valley, left panel in figure \ref{fig2}(c,d) is opposite to the level arrangement in figure \ref{fig2}(a,b). There is no inversion of electronic levels for the electron in the presence of valence hole in the $-K$ valley. Hence, the T$_T$ trion is a triplet and the lowest energy inter-valley trion is dark. The T$_S$ trion is a singlet trion, as the T$_1$ line in MoS$_2$. The observation of the doublet structure of bright high-energy trions (T$_S$, T$_T$) in WSe$_2$ and WS$_2$ monolayers results from the fact that in tungsten based monolayers the optically active exciton (X) is associated with the top spin-split valence subband (VB) and the upper spin-split conduction band subband (CB). Hence, for low two-dimensional electron gas concentration, as the electron Fermi level is positioned between the spin-split conduction bands, the triplet trion (figure \ref{fig2}(c)) comprises two electrons from different valleys \cite{Yu-Yao-2015}, whereas the singlet trion must involve two electrons from the same valley (figure \ref{fig2}(d)). Additionally, in the simplest case of only one charged complex, the T$_1$ trion in MoSe$_2$ (figure \ref{fig1}(b)) is formed from the optically active exciton (X) associated with the top spin-split valence band (VB) subband and the lower spin-split conduction band (CB) subband, see figure \ref{fig2}(e). Accordingly, for the low 2DEG concentration and Fermi level positioned close to the lower spin-split conduction band, the singlet trion should involve two electrons from the different valleys forming the inter-valley spin singlet trion (T$_S$) with two electrons located in the lower conduction bands.

The assignment of T$_2$ (figure \ref{fig2}(a)) and T$_1$ (figure \ref{fig2}(b)) emission features in MoS$_2$ to the inter-valley and intra-valley bright singlet trions instead of singlet-triplet trions is related to the relatively small, compared with WS$_2$ or WSe$_2$ monolayers, spin-splitting and Fermi level position in the conduction band. We note that the inter-valley triplet trion in such band arrangement is not only less probable due to Fermi level position, but was also predicted to be unbound due to exchange interaction \cite{Tempelaar-Berkelbach-2019}.

\begin{figure*}
    \centering
    \includegraphics[width=0.95\textwidth]{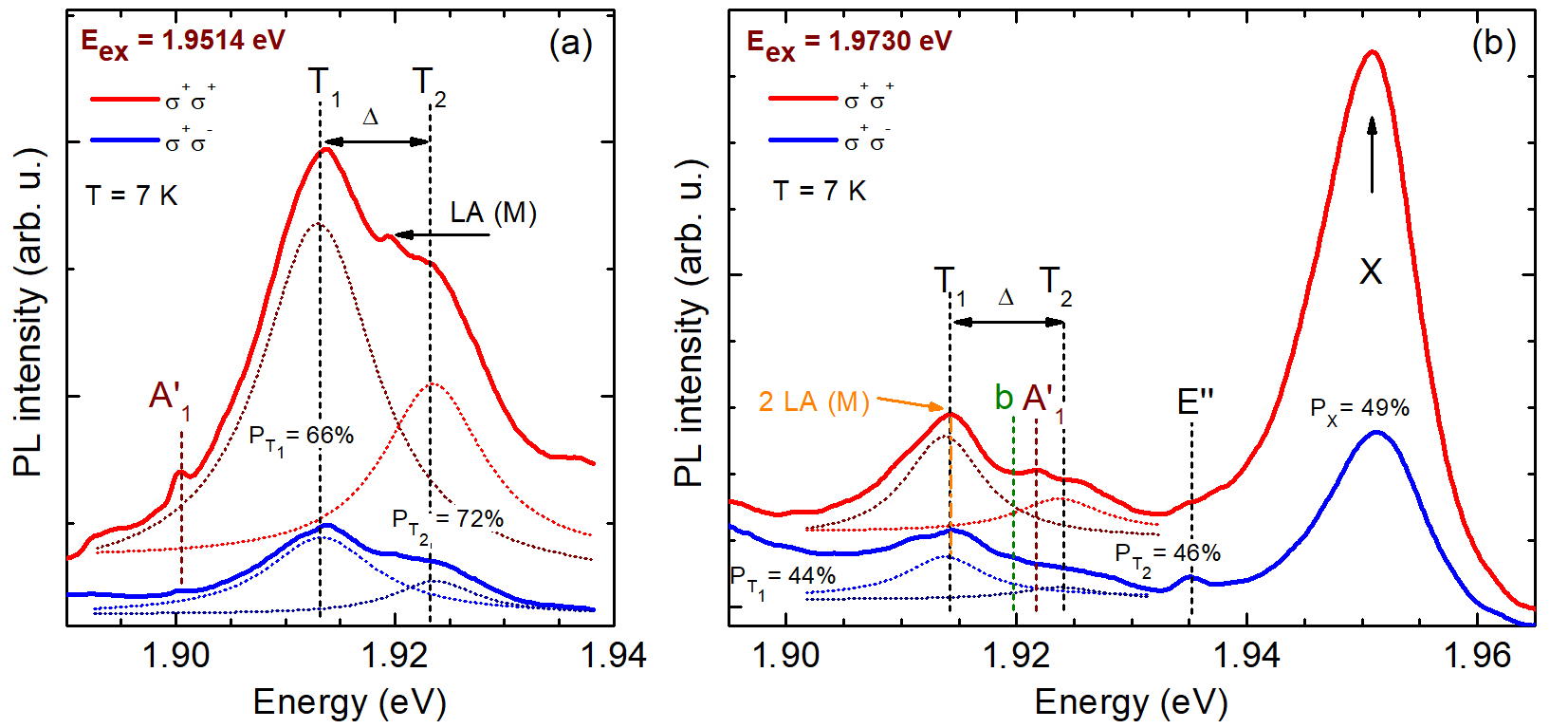}
    \caption{(a) The polarization-resolved PL spectra of MoS$_2$ excited strictly resonantly at the neutral exciton energy of $1.9514$ eV. (b) The resonant circularly-polarized PL spectra for $1.9730$ eV excitation, revealing $E^{''}$, $A_{1}^{'}$, $b$ and $2 LA$ ($M$) Raman features. } 
		\label{fig3}
\end{figure*}

\section{Polarization-resolved PL excitation of trions in monolayer MoS$_2$} To firmly establish the singlet nature of the T$_1$ and T$_2$ transitions observed in the PL spectra of monolayer MoS$_2$, we perform polarization-resolved and excitation energy-dependent PL measurements at $T = 7$ K. Figure \ref{fig3} shows the examples of polarization-resolved PL spectra excited strictly resonantly at the neutral exciton energy of 1.9514 eV (figure \ref{fig3}(a)) and near-resonantly at the energy of 1.9730 eV (figure \ref{fig3}(b)). We observe also the Raman features, which shift as a function of the excitation energy and are attributed to first- and second-order scattering processes, such as $A_{1}^{'}$, $E^{'}$ modes and $b$, $2 LA$ ($M$) or $LA$ ($M$) bands, respectively \cite{Molas-2019, Kutrowska-Girzycka-Bryja-2018}. The presence of the $LA(M)$ band in the Raman scattering spectrum is likely due to defects or disorder in the sample which localize phonons \cite{Molas-2019}. Furthermore, we analyze the optical orientation of trions, i.e. the helicity of the outgoing light with respect to the helicity of their emission. Based on the fitting of the trion contributions to the PL spectrum excited at the neutral exciton energy X (figure \ref{fig3}(a)) with a combination of two Lorentzian curves, we determine the degree of helicity preservation of the emitted light with respect to exciting light defined as $P = (I_{\sigma^{+}\sigma^{+}} \ensuremath{-} I_{\sigma^{+}\sigma^{-}})/(I_{\sigma^{+}\sigma^{+}} + I_{\sigma^{+}\sigma^{-}})$ for each trion component, where $\sigma^{+}\sigma^{+}$ and $\sigma^{+}\sigma^{-}$ indicate the co-circular and cross-circular configurations, respectively. The T$_1$ line displays the P$_{T_{1}}$ value of almost $66\%$, while the T$_2$ line has a slightly higher P$_{T_{2}}$ of about $72\%$. For the excitation energy equal to 1.9730 eV (slightly higher than the X energy) we obtain P values of $44\%$, $46\%$ and $49\%$ for T$_1$, T$_2$ and X, respectively (figure \ref{fig3}(b)). In monolayer TMDs, the large exciton valley polarization obtained in the steady-state PL experiments results from the competition between the valley depolarization time ($\sim 1$ ps) and the ultra-fast exciton population relaxation time ($\sim 100$ - $200$ fs) \cite{Moody-Li-2015, Moody-Xu-2016}. However, the trions exhibit an extended population relaxation time of tens of picoseconds. It has been established in a recent work \cite{Singh-Li-2016} that the most efficient scattering mechanism responsible for the trion valley depolarization in WSe$_2$ is due to scattering of an electron-hole pair between valleys. This process is mediated by the electron-hole exchange interaction towards an energetically favorable trion state. In contrast to tungsten based TMDs, where the preservation of helicity of the excitation light in the emission spectrum of T$_S$ and T$_T$ trions is significantly different, e.g. about $16\%$ and $50\%$ in steady-state PL measurements of monolayer WS$_2$ \cite{Vaclavkova-Molas-2018}, in monolayer MoS$_2$ both values are comparable. This supports the interpretation that both trion lines originate from singlet states.

\section{Trion splitting as a measure of correlations} Having established the singlet nature of the T$_1$ and T$_2$ trions, we comment on processes determining their energy splitting. The difference in the total energy of the T$_1$ and T$_2$ configurations $E_{T_{1}} - E_{T_{2}} = \Delta_{SOC} + V^{*}$, can be written as a sum of conduction band spin splitting $\Delta_{SOC}$ and electron-electron interactions, $V^{*}$. $V^{*}$ reflects a subtle imbalance between electron-hole and electron-electron interactions inside and between valleys. A simple analysis leads to the expression

\begin{equation}
\eqalign{
V^{*} =& \left[V^{D}(e_{1},+K; e_{2},+K) - V^{D}(h_{1},+K; e_{2},+K)\right] +\cr
&\left[V^{D}(e_{1},+K;e_{2},-K) - V^{D}(h_{1},+K;e_{2},-K\right],}
\end{equation}
where $e_{1,2}$ ($h_{1}$) describe electrons (hole) forming trion (see figure \ref{fig2}(a)) and $V^{D}$ is the total direct interaction energy between particles. We see that $V^{*}$ is the sum of two contributions, a difference between the electron-electron repulsion and electron-hole attraction in the same valley and between valleys. Interestingly, a similar analysis of the singlet-triplet trion splitting in tungsten-based materials leads to $E_{T_{S}} - E_{T_{T}}=V^{X}(e_{1},+K;e_{2},-K) + V^{*}$. It suggests that the energy of the singlet-singlet splitting is a new measure for electronic correlations in 2D crystals with small spin-orbit splitting in conduction bands.

\section{Zeeman g-factor in monolayer MoS$_2$} 
\begin{figure}[h]
    \centering
    \includegraphics[width=0.45\textwidth]{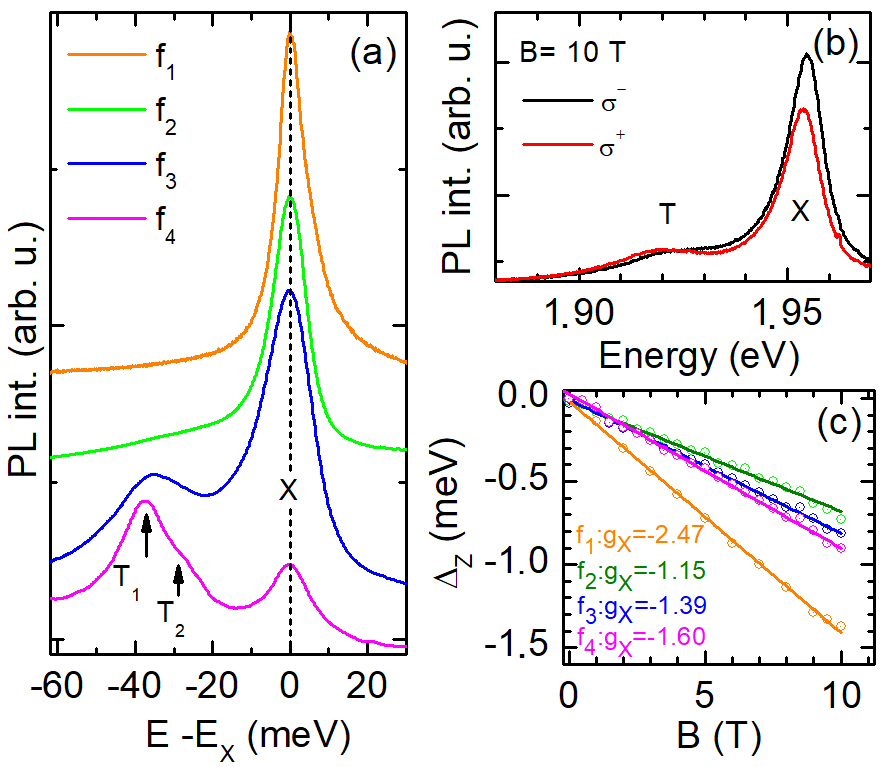}
    \caption{(a) The comparison of low temperature (7 K) PL spectra of diﬀerent MoS$_2$ monolayers: hBN-encapsulated - f$_1$, and three MoS$_2$/hBN structuress with diﬀerent thicknesses of hBN layers: f$_2$-100 nm, f$_3$-130 nm, f$_4$-250 nm, respectively. (b) The circularly-polarized PL spectrum for f$_3$ sample at $B=10$ T. (c) Exciton Zeeman splitting. } 
		\label{fig4n}
\end{figure}

The relative PL intensity of T$_1$ and T$_{2}$ features significantly depends on the quality of the sample and doping level. Figure \ref{fig4n}(a) compares low temperature (7 K) PL spectra of four different MoS$_2$ monolayers: (i) one hBN-encapsulated (hBN/MoS$_2$/hBN)- f$_1$ and (ii) three monolayers with only down hBN substrate (MoS$_2$/hBN) with different hBN thickness: f$_2$-100 nm, f$_3$-130 nm, f$_4$-250 nm, respectively. Similarly to our previous studies on WS$_2$/hBN/SiO$_2$ structures \cite{Jadczak-2019}, an extra hBN layer used between the flake and SiO$_2$/Si substrate acts as a buffer layer and changes the doping level in the monolayer system. Accordingly, the presented PL spectra reveal relatively different 2D electron gas (2DEG) concentrations, which is qualitatively estimated by the trion to exciton emission intensity ratio $(T/X)$ \cite{Jadczak-2019, Jadczak-Bryja-2017}. At very low 2DEG concentration, foreseen for the hBN-encapsulated sample (f$_1$), the X line solely dominates the PL spectrum (Fig. \ref{fig4n}(a), orange line), whereas both T lines are not detectable. Also, for one of the MoS$_2$/hBN samples (f$_2$, green line), for which the neutral exciton is very intensive; however, below its energy position the PL spectrum exhibits a dim and broad trion shoulder. Moreover, for both the hBN-encapsulated (f$_1$) and MoS$_2$/hBN (f$_2$) samples the linewidth (FWHM) is most narrow and reaches about 5 meV at 7 K. At a slightly higher 2DEG concentration (sample f$_3$), the emission intensity of broaden T transitions increases. Nevertheless, the double trion structure becomes well resolved only for the highest electron doping (f$_4$, magenta line), as the trion emission intensity exceeds that of a neutral exciton.  Our observations are consistent with recent results concerning an optical susceptibility measurement of a gated MoS$_2$ device \cite{Roch-Warburton-2019}.

Let us now discuss the magneto-optical response observed in the PL spectra at $T = 5$ K. Figure \ref{fig4n}(b) presents the typical, polarization-resolved PL spectrum for MoS$_2$/hBN at $B=10$ T (f$_3$). Using non-resonant (2.331 eV) and linearly polarized laser excitation we find that, for the applied magnetic fields and all the samples, the trion features are not sufficiently resolved in the PL spectra to estimate their effective g-factors. Accordingly, we focus only on the 2DEG dependence of the exciton g-factors.

\begin{figure}[h]
    \centering
    \includegraphics[width=0.45\textwidth]{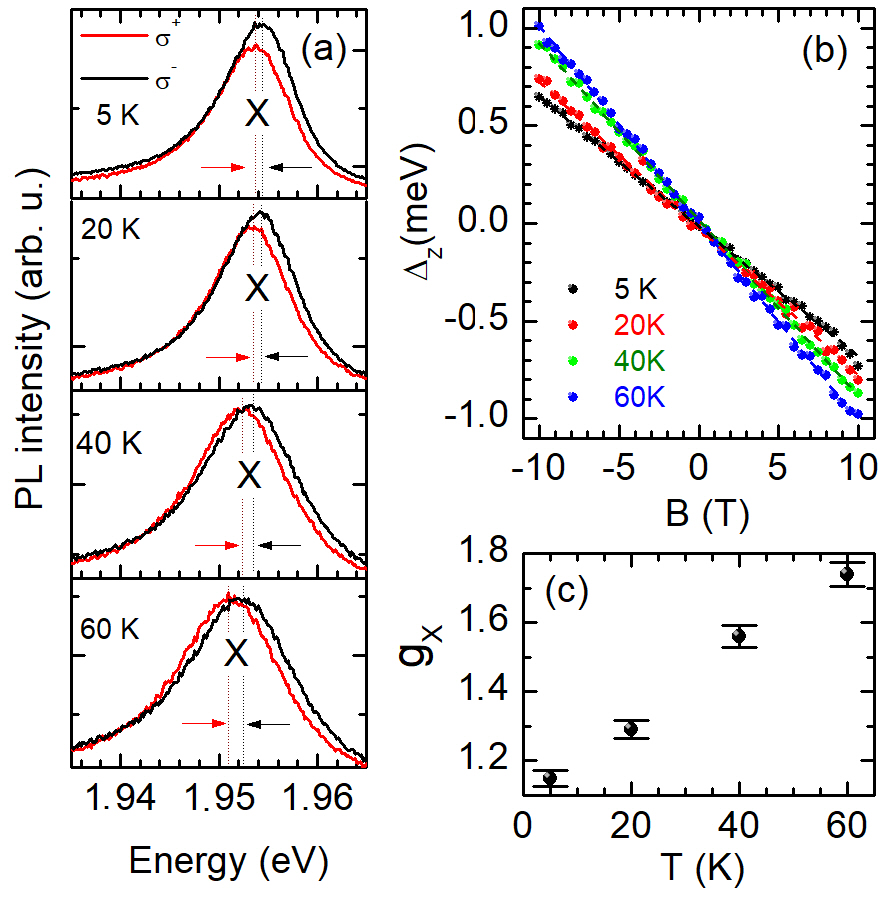}
    \caption{(a) Typical PL spectra of MoS$_{2}$/hBN structure ($f_2$) recorded at 5 K, 20 K, 40 and 60 K. (b) Temperature dependence of exciton Zeeman splitting. (c) Temperature evolution of the effective exciton g-factor.  } 
		\label{fig5}
\end{figure}

In figure \ref{fig4n}(c) we extracted the Zeeman splitting $\Delta_{Z}$ of the neutral exciton X for all the samples; it is defined as the shift between the $\sigma^{+}$ and $\sigma^{-}$ polarized components of the PL: $\Delta_{Z} =E_{\sigma^{+}}-E_{\sigma^{-}}=g_X\mu_{B}B$. This quantity depends linearly on the magnetic field and our measurement suggest the exciton g-factor of $g_{X} =-1.15\pm 0.01$, $g_{X} =-1.39\pm 0.01$, $g_{X} =-1.60\pm 0.01$ and $g_{X} =-2.45\pm 0.01$ for samples f$_2$, f$_3$, f$_4$ and f$_1$, respectively. Interestingly, for all the MoS$_2$/hBN structures without a cap layer the exciton g-factor increases with rising thickness of the hBN layer: f$_2$-100 nm, f$_3$-130 nm, f$_4$-250 nm, respectively. For the high-quality hBN-encapsulated monolayer MoS$_2$ (f$_1$) the exciton g-factor is the highest. 
Furthermore, we probed the temperature dependence of the exciton effective g-factor in the MoS$_2$/hBN structure (f$_2$), whose emission spectra are dominated only by a sharp and relatively narrow X line. Figure \ref{fig5}(a) compares typical PL spectra of the MoS$_{2}$/hBN structure with relatively low doping, measured at different temperatures ($5$ K, $20$ K, $40$ K, $60$ K) and a magnetic field of $B = 10$ T. The exciton Zeeman splitting at four different temperatures is shown in figure \ref{fig5}(b). As seen in figure \ref{fig5}(c), the absolute value of the exciton g-factor increases from 1.15 to 1.74 for increasing temperature from 5 K to 60 K. This increase of the exciton g-factor by about $34\%$ is related to the temperature broadening of the X emission line (see figure \ref{fig5}(a)). However, it may also result from different thermal distributions of electrons in the spin-split subbands. Interestingly, in the high-quality hBN-encapsulated monolayer MoS$_2$ the temperature effect on the exciton g-factor is not observable.

Our results are consistent with previously reported studies of high-quality hBN-encapsulated MoS$_{2}$ monolayers with very narrow exciton linewidths (2-4 meV), showing exciton g-factor of $g_{X}= -1.7$ \cite{Cadiz-Urbaszek-2017} in magneto-PL or $g_{X}= -2.9$ \cite{Goryca-Crooker-2019} in magneto-transmission experiments. It is worthwhile to mention that the Zeeman splitting, and hence g-factors, seem to depend on the overall optical quality of the sample (doping level), which in turn can be reflected in the linewidth of the PL or RC transitions. However, in comparison to other monolayer TMDs \cite{Goryca-Crooker-2019,Koperski-Potemski-2019}, the small exciton Zeeman splitting in the monolayer MoS$_2$ may arise from the interaction with close in energy, spin- and valley-forbidden, dark excitons. Also, as we have shown, this interaction is tuned by different dielectric environments of MoS$_2$. However, it needs further theoretical investigations.

\section{Conclusions}
In summary, we present results of optical experiments and theoretical analysis on the high-quality single-layer MoS$_{2}$ which reveal the fine structure of charged excitons, i.e., trions. In the emission spectra we resolve and identify two trion peaks, T$_{1}$ and T$_{2}$, resembling the pair of singlet and triplet trion peaks (T$_S$ and T$_{T}$) in tungsten-based materials. In polarization-dependent photoluminescence measurements we identify these peaks as intra- and inter-valley singlet trions due to the trion fine structure distinct from that already known in bright and dark 2D materials with large conduction-band splitting induced by the spin-orbit coupling. With additional support from theory we claim that the existence of these singlet trions combined with an anomalous excitonic g-factor together suggest that the monolayer MoS$_{2}$ has a dark excitonic ground state, despite having a "bright" single-particle arrangement of spin-polarized conduction band states.

\ack
J.J. J.K.-G. and L.B. acknowledge support by the Polish NCN Grant "Beethoven 2" No. 2016/23/G/ST3/04114. J.D. and J.J.S. acknowledge support by the German DFG "Beethoven 2" Grant No. DE 2206/2-1. M.B. and P.H. thank P. L. Lo, S. J. Cheng, NCTU Taiwan and L. Szulakowska, uOttawa for discussions. M.B. and P.H. acknowledge support from NSERC Discovery and QC2DM Strategic Project grants as well as uOttawa Research Chair in Quantum Theory of Materials, Nanostructures and Devices. A.W. and M.B. acknowledges financial support from National Science Center (NCN), Poland, grant Maestro No. 2014/14/A/ST3/00654. Computing resources from Compute Canada and Wroclaw Center for Networking and Supercomputing are gratefully acknowledged. T.K. and P. K. acknowledge support by the ATOMOPTO project carried out within the TEAM programme of the Foundation for Polish Science co-financed by the European Union under the European Regional Development Fund.

\section*{References}
\bibliography{IOP_Trions} 
\bibliographystyle{unsrt}

\end{document}